\documentclass[authoryear,12pt]{article}
\usepackage[english]{babel}
\usepackage{url,amsfonts,epsfig}
\usepackage{amsfonts}
\usepackage{stmaryrd}
\usepackage{mathrsfs}
\usepackage{amsmath}
\usepackage{enumerate}
\usepackage{mathrsfs}
\usepackage{mathtools}
\usepackage{graphicx}
\usepackage{amsthm}
\usepackage{natbib}
\usepackage{subfigure}
\usepackage{tikz}
\usepackage{vmargin}
\usepackage{amssymb}
\usepackage{lscape}
\setpapersize{A4}
\newcommand{\diag}{\ensuremath{\mathrm{diag}}}

\newcommand{\cov}{\ensuremath{\mathrm{Cov}}}

\setmarginsrb{23mm}{23mm}{23mm}{30mm}{0pt}{0mm}{0pt}{0mm}
\pdfoutput=1
\begin{document}
\title{Combining data from multiple spatially referenced prevalence surveys using generalized linear geostatistical models}
\author{Emanuele Giorgi$^{1,2}$, Sanie S. S. Sesay$^{3,4}$, Dianne J. Terlouw$^{3,4}$  \\ and Peter J. Diggle$^{1,2}$}

\maketitle
\begin{abstract}
Data from multiple prevalence surveys  can provide information on common parameters of interest, which can therefore be estimated more precisely in a joint analysis than by separate analyses of the data from each survey. However,  fitting a single model to the combined data from multiple surveys is inadvisable without testing the implicit assumption that all of the surveys are directed at the same inferential target.  In this paper we propose a multivariate generalized linear geostatistical model that accommodates two sources of heterogeneity across surveys so as to correct
 for spatially structured bias in non-randomised surveys and to
allow for temporal variation in the underlying prevalence surface between consecutive survey-periods.

 We describe a
Monte Carlo maximum likelihood procedure for parameter estimation, and show through
simulation experiments how accounting for the different sources 
of heterogeneity among surveys in a joint model leads to more precise inferences. 
We describe an application to multiple surveys of
malaria prevalence conducted in Chikhwawa District, Southern Malawi, and discuss how this approach could
inform hybrid sampling strategies that combine data from randomised and 
non-randomised surveys so as to make the most efficient use of all available data.  \\
\text{ } 
\end{abstract}

\begin{small}
1. Medical School, Lancaster University, Lancaster, UK\\
2. Institute of Infection and Global Health, University of Liverpool, Liverpool, UK\\
3. Malawi-Liverpool-Wellcome Trust Clinical Research Programme, Blantyre, Malawi.\\
4. Liverpool School of Tropical Medicine, Liverpool, UK
\end{small}

\pagebreak

\section{Introduction}
\label{sec:intr}
In studies of spatial variation in disease prevalence, 
 it is often necessary to 
combine information from multiple prevalence surveys. This is particularly the case
in  low-resource settings, where disease registries 
typically do not exist. A methodological challenge in these circumstances
is that survey designs are
severely constrained by cost constraints.
The available surveys
may therefore
be of variable quality and/or conducted at different times. In this paper, we 
propose a class of generalized linear geostatistical models (GLGMs) to address two
specific issues.The first
is variation in quality, for example between randomised and non-randomised surveys, in which case our proposed methodology  assumes that 
at least one of the surveys provides an
unbiased ``gold-standard''. The second is variation in the underlying prevalence
when surveys are conducted at different times. In this case, by modelling the underlying prevalence over time we
are able to use data collected at all times to estimate the underlying prevalence surface at the specific time of interest,
typically the time of the most recent survey.

Methods for the combined analysis of data from multiple surveys
have previously used meta-analysis and small area statistics
approaches; see \citet*{statmatch2001}, \citet*{cancer2005}, \citet*{multframe2006}
 and \citet*{biasmod2009}. More recently,
\citet*{combsurv2011} used
 Bayesian hierarchical models to combine smoking prevalence estimates 
from multiple surveys. They noted that commercial surveys are often
ignored in constructing official estimates
because of poor information about the sampling designs used, but argued that these surveys
can nevertheless
provide useful additional information because they are
more frequently updated than official surveys.

\citet*{raghu2007} noted the potential benefits that might 
accrue from spatial modelling of multiple survey
data, but to the best of our knowledge,
explicit spatial modelling of biases and/or temporal variation
in the outcome of interest has not previously been addressed, except in a few
specific applications. For example, \citet*{wanji2012}
established a logit-linear calibration relationship between 
estimates of {\it Loa loa} prevalence
in part of equatorial Africa based on 
two different methods,
finger-prick
blood sampling and  a short questionnaire instrument.
\citet*{crainiceanu2008}  incorporated this 
calibration relationship
into
a bivariate geostatistical model for the two corresponding prevalence maps.

As discussed in \citet*{biasmod2009}, if information from multiple surveys is to be
combined, it is important to understand
the limitations of their designs in order to take account of potential
biases in the associated estimates of prevalence. As a minimal condition, the
study subjects in each survey
should be drawn from the same target population.
One potential source of bias is that
some members of the target population may be less likely than others
to be included. 
Convenience samples 
provide an example of this. In resource-poor settings, the relatively low cost of 
convenience sampling is tempting, but its potential to produce
 biased estimates is clear. In a non-spatial context,
  \citet*{hedt2011} propose a hybrid prevalence estimator 
that combines information from randomised and convenience surveys.
They demonstrate that, with suitable adjustment for the bias, their hybrid estimator
can give better prevalence estimates than would be obtained by using only the data 
from the randomised surveys.

A second source of heterogeneity amongst multiple prevalence
surveys is temporal variation in prevalence. When spatially referenced
prevalence surveys are repeated over time it
is usually of interest to estimate changes in prevalence over time.
When the outcomes from consecutive surveys are correlated, there is also a potential gain in
efficiency if comparisons are made through the use of a joint model. This is especially
advantageous when the surveys do not use the same set of sampling locations, because a joint 
analysis can then exploit both the temporal and spatial correlation structure of the
combined data.

In Section \ref{sec:mod_form} of the paper we propose a class of generalised linear
geostatistical models (GLGMs) for the combined analysis
of data from multiple prevalence surveys. The model allows both for biased sampling and
temporal variation in prevalence provided that one of the surveys delivers
unbiased ``gold-standard'' estimates of prevalence.   In Section \ref{sec:inference}
we describe the methods that we use to fit the model.
In Section \ref{sec:simstudy}
we report the results 
of simulation experiments that illustrate how a joint model leads to 
gains in efficiency of estimation and spatial prediction.
In Section \ref{sec:application} we describe an application  to malaria prevalence data from three surveys conducted
in Chikhwawa District, Southern Malawi. 
Section \ref{sec:discussion} is a concluding discussion.
All computations for the paper were run on the High End Computing Cluster
at Lancaster University, using the R software environment \citep*{rsoftware}.

\section{A multivariate generalized linear geostatistical model}
\label{sec:mod_form}

The ingredients of a univariate
GLGM are the following. Random variables $Y_j$ and explanatory variables
$d_j$
are associated with sampling locations
$x_j$ in a region of interest $A \subseteq \mathbb{R}^2$.  Each $d_j$ is a vector of
length $p \geq 1$.
Conditional on the realisation of a zero-mean
latent Gaussian process $S(x)$ and a set of mutually independent zero-mean
latent Gaussian variables $Z_j$, the $Y_j$ follow a classical
generalized linear model \citep*{GLM1989}, hence:
\begin{itemize}
\item[(i)] the $Y_j$ are mutually independent conditional on the $S(x_j)$
and $Z_j$, with conditional
expectations $\mu_j = m_j g^{-1}(\eta_j)$, where $m_j$ is a known
scalar and $g(\cdot)$ a known {\it link function};
\item[(ii)] $\eta_j = d_j^\top \beta + S(x_j) + Z_j$;
\item[(iii)] the conditional distribution of the $Y_j$ falls within the exponential family.
\end{itemize}
In the remainder of the paper, we assume that the conditional distributions in (iii)
are binomial, with the $y_j$ representing the number of positives amongst $m_j$ individuals
sampled at location $x_j$. We also adopt the  standard logistic link function,
$g(\mu) = \log \{\mu/(1-\mu)\}$, but other link functions could also be used. 
We specify the Gaussian process $S(x)$ to have
covariance function $\cov\{S(x),S(x')\}=\sigma^2\rho(x,x';\phi)$, 
and the mutually independent $Z_j$ to have variance $\tau^2$.
The $Z_j$
have a dual interpretation as either non-spatial extra-binomial
variation or spatial variation at scales smaller  than the
smallest distance between sampling locations; the two interpretations
can only be disentangled unambiguously
if repeated measurements are taken at coincident locations. Finally, we write 
$d_j = d(x_j)$  to
emphasise its spatial context. 
\begin{figure}[ht]
\vspace{-0.2in}
\begin{center}
\includegraphics[scale=0.6]{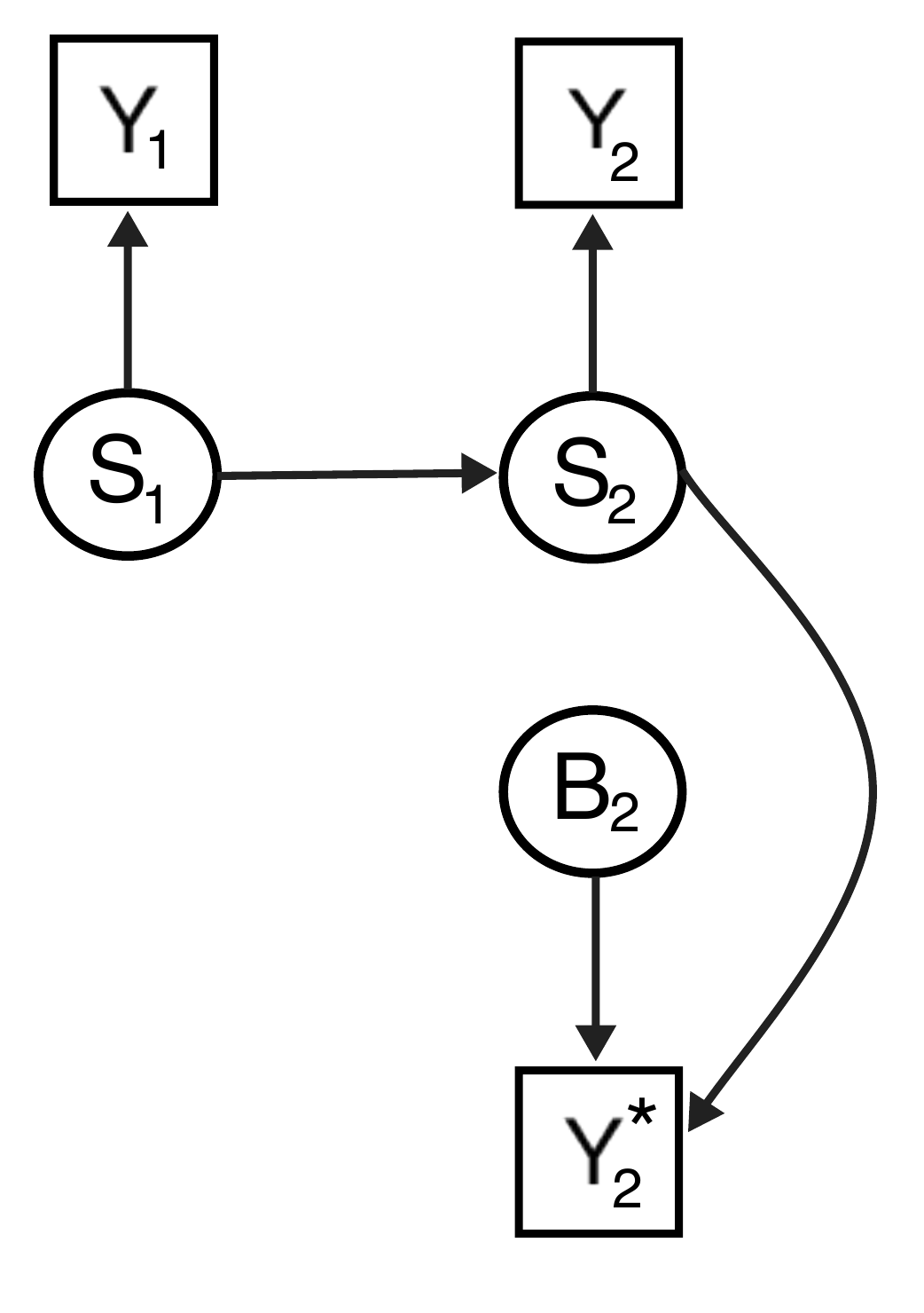}
\caption{Representation of the multivariate generalized linear geostatistical model \eqref{eq:GLGM} as a directed acyclic graph; $S_1$ and $S_2$ represent prevalences at
times $t_1$ and $t_2>t_1$; $B_2$ represents bias; $Y_1$, $Y_2$ and $Y_{2}^*$  
are observed prevalences from unbiased surveys at times $t_1$ and $t_2$, and 
a biased survey at time $t_2$, respectively. The target for predictive inference is
$S_2$. \label{fig:graph}}
\end{center}
\end{figure}

To extend the  model to accommodate multiple surveys taken at possibly different times, some of which may be biased,
let $i=1,\dots,r$ denote the index of the survey and $x_{ij}: j = 1,\ldots,n_{i}$
the corresponding set of sample locations. We replace the single process $S(x)$ by
a set of $r$ processes $S_i(x)$ which relate to the true prevalence at different
times. We assume that at least one of the surveys is known to be
unbiased, define ${\cal B}$ to be the
index set of the potentially biased surveys
and introduce an additional set of  latent 
Gaussian processes $B_i(x): i \in {\cal B}$ 
to represent the spatially
varying  biases. Finally, we assume that data from different surveys
are generated by conditionally 
independent univariate GLGMs, with link functions
\begin{eqnarray}
\label{eq:GLGM}
g_i(\mu_{ij}/m_{ij}) &=&  \eta_{ij} = d(x_{ij})^\top \beta_1 + S_{i}(x_{ij}) + 
        I(i \in {\cal B})[B_i(x_{ij}) + d(x_{ij})^\top \beta_i]+  Z_{ij},\nonumber \\
            && j=1,\ldots,n_{i}; i=1,\ldots,r.
\end{eqnarray}
On the right-hand-side of (\ref{eq:GLGM}), we assume that the marginal properties of each
$S_i(x)$  are the same 
as previously specified for $S(x)$, and add a set of cross-covariance
functions, $\cov\{S_{i}(x),S_{i'}(x')\}=\sigma^2 \alpha_{ii'} \rho(x,x';\phi)$,
where $-1 < \alpha_{ii'} < 1$. The parameters $\alpha_{ii'}$
capture the temporal correlation between
the true prevalence surfaces at different times, hence if surveys $i$ and $i^\prime$
are taken at the same time, $S_i(x) = S_{i^\prime}(x)$ for all $x$ 
and $\alpha_{ii'}=1$. Note that if $r>2$, some combinations
of $\alpha_{ii'}$ result in a non-positive-definite variance matrix. If $r$ is small, this can be handled 
by setting the likelihood to zero for all such combinations. When $r$ is large the issue can be
avoided 
by imposing a spatio-temporally continuous parametric structure. This has the incidental benefit of  making the model more
parsimonious. One such example would be an exponentially decaying cross-covariance structure 
 with $\alpha_{ii'} = \exp\{-|t_i - t_{i'}|/\psi\}$, where $t_i$ is the time at which the
$i$th survey is taken. 

The processes $B_i(x)$ in (\ref{eq:GLGM})
are assumed to be independent, with
zero mean and covariance functions 
$\cov\{B_i(x),B_i(x')\}=\nu_i^2\rho(x,x';\delta_i)$.

Finally, the random variables $Z_{ij}$ are again assumed to be mutually independent and Normally distributed
with common mean 0 and variances $\tau_i^2$.

As already noted, when all surveys are taken at the same time, $S_i(x)=S_1(x)$ for all $i$, which formally
corresponds to $\alpha_{ii^\prime}=1$ for all $(i,i^\prime)$. When 
all surveys are unbiased but are taken at different times, ${\cal B}$ is the empty set and 
the terms $[B_i(x_{ij}) + d(x_{ij})^\top \beta_i]$ in (\ref{eq:GLGM}) are omitted; formally, this
 corresponds to $\nu_i^2=0:i=2,\ldots,r$. If it is appropriate to use different 
explanatory variables to model the true prevalence and the bias, this is 
accommodated by setting some elements of the $\beta_i$ to zero. 
The dependence structure of the model is illustrated by the directed acyclic graph in Figure \ref{fig:graph} for the special case of two gold-standard surveys conducted at two different times and a biased survey at the second time period. This scenario corresponds to case study analysed in Section \ref{sec:application}, where the aim is predictive inference for $S_2(x)$. In this case, the potential gains in efficiency
by jointly modelling the  data from all  three surveys stem from the direct
links between  $S_2$ and both $Y_2$ and $Y_2^*$ and the indirect link 
between $S_2$ and $Y_1$ via $S_1$.

\section{Inference}
\label{sec:inference}

In this section, we focus on the case $r=2$.
The generalization to more than two surveys 
is straightforward. We set $B_1(x)=0$,  write $B(x)$ in place of $B_2(x)$ and 
write the parameters of this bivariate version of (\ref{eq:GLGM}) 
as $\beta=(\beta_1,\beta_2)$ and
 $\theta = (\sigma^2, \nu^2,\tau_1^2,\tau_2^2,\phi,\delta,\alpha)$.

\subsection{Likelihood}
\label{subsec:lik}

Let $y_{i} = (y_{i1},\ldots,y_{in_{i}})^\top$ denote the outcome data from surveys
$i=1,2$ and let $D_i$ be the $n_i$ by $p$ matrix whose $j$th row contains the
values $d(x_{ij}) = (d_1(x_{ij}),\ldots,d_p(x_{ij}))^\top$. Similarly, let
$T_i$ denote the vector of the $n_i$ values of
the linear predictor for survey $i$,
hence $T_i = D_i \{\beta_1 + I(i=2)\beta_2\} + W_i$, 
where $W_i = (W_{i1},\ldots,W_{in_i})^\top$
and
\begin{equation}
W_{ij} = S_{i}(x_{ij}) + I(i=2)B(x_{ij}) + Z_{ij}.
\label{eq:random_part}
\end{equation}
Now, let $T$ denote
the $(n_1+n_2)$-element vector $T = (T_1^\top,T_2^\top)^\top$ and 
$D$ the $(n_1+n_2)$ by $2p$ matrix,
\begin{equation}
D = \left[
\begin{array}{ll}
D_1 & 0 \\
D_2 & D_2 
\end{array}
\right].
\label{eq:mean}
\end{equation}
Also, write $R_{ii'}(\phi)$ for the $n_i$ by $n_{i'}$ matrix with 
$(h,k)$th element $\rho(x_{ih},x_{i^\prime k};\phi)$ and 
$R_b(\delta)$ for the $n_2$ by $n_2$ matrix with 
$(h,k)$th element $\rho(x_{2h},x_{2k};\delta)$.
Then, 
\begin{equation}
T \sim {\rm MVN}(D \beta, V(\theta))
\label{eq:MVN}
\end{equation}
where
\begin{equation}
V(\theta) = 
\left[
\begin{array}{ll}
\sigma^2 R_{11}(\phi) + \tau_1^2 I & \sigma^2 \alpha R_{12}(\phi) \\
\sigma^2 \alpha R_{21}(\phi)       & \sigma^2 R_{22}(\phi) + \nu^2 R_b(\delta) + \tau_2^2 I
\end{array}
\right].
\label{eq:covmat}
\end{equation}
The conditional distribution of $Y$ given $T=t$ is a product of independent binomial
probability mass functions. We write this as
\begin{equation}
f(y|t) = \prod_{i=1}^{2}\prod_{j=1}^{n_{i}}f(y_{ij}|t_{ij}).
\label{eq:conditional}
\end{equation}
Combining (\ref{eq:mean}), (\ref{eq:MVN}), (\ref{eq:covmat}) and (\ref{eq:conditional}) then
gives the likelihood function as the  high-dimensional integral
\begin{equation}
\label{eq:likelihood}
L(\beta,\theta) = \int  h(t; D\beta, V(\theta))f(y|t) \: dt,
\end{equation}
where $h(\cdot|\mu,V)$ is the density function of a multivariate Normal 
distribution with mean $\mu$ and covariance matrix $V$.

\subsection{Conditional simulation}
\label{subsec:simre}

We propose to use Monte Carlo methods to evaluate the high-dimensional
integral in (\ref{eq:likelihood}). These methods require us to  simulate 
from the conditional distribution of the spatial random effect 
$T$ given the data $Y=y$. Using Bayes' formula,
this conditional density is
\begin{equation}
\label{eq:conditional_dens}
\pi(t |y) \propto h(t|D \beta, V(\theta))f(y|t).
\end{equation}
To simulate from (\ref{eq:conditional_dens}),
\citet*{mcmc2006} propose a Langevin-Hastings (LH) Markov chain 
Monte Carlo (MCMC) algorithm. This operates by updating a 
linear transformation of $T$, chosen to make the components of $T|y$ approximately 
independent.
\citet*{mcmc2006} 
use a Gaussian approximation to the distribution of $T|y$, with
mean $D \beta$ and covariance matrix
\begin{equation}
\label{eq:sigmatilde}
\tilde{V} = \{V(\theta) +\Lambda(\hat{t})\}^{-1}.
\end{equation}
In (\ref{eq:sigmatilde}), $\Lambda(t)$ is a diagonal matrix with entries 
$-\partial^2/\partial t_{i}^2 \log f(y|t)$ and 
$\hat{t}$ is a typical value of $T$ such as  
the mode of $f(y|t)$.  For 
the binomial model with logistic link, this gives
 $\Lambda(\hat{t}) = \diag\{y_{i}(1-y_{i}/m_{i})\}$.
\citet*{mcmc2006} demonstrate that updating the centred random variable 
$
\tilde{T} = \tilde{V}^{-1/2}(T-D \beta)
$
gives better mixing and convergence properties than the analogous
MCMC algorithms based on either $T$ or on $\bar{T} = V^{-1/2}(T-D \beta)$, as suggested by \citet*{waag2002}.

\subsection{Monte Carlo Maximum Likelihood: estimation and spatial prediction}
\label{subsec:mmle}

The Monte Carlo Maximum Likelihood (MCML) method \citep*{geyer1992, geyer1994, geyer1996, geyer1999}
uses conditional simulations of $T$ given $Y$ to obtain a computationally efficient
approximation to the intractable 
likelihood function. 
From \eqref{eq:likelihood}, the likelihood function can be written as
\begin{eqnarray}
\label{eq:likelihood2}
L({\beta},{\theta}) &=& \int  h({t}|D{\beta}, V({\theta}))f({y}| {t}) \: d{t} 
                                             = \int  \frac{h({t}|D{\beta}, V({\theta}))f({y}| {t})}
                                                                {\tilde{f}({y},{t})}\tilde{f}({y},{t})\: d{t} \nonumber \\
                             & \propto &  \int \frac{h({t}|D{\beta}, V({\theta}))f({y}| {t})}
                                                          {\tilde{h}({t})f({y}| {t})}
                                                          \tilde{f}({t} | {y})\: d{t}  
                                                          = 
                                                          E_{\tilde{f}}\left[\frac{h({t}|
                                                          D{\beta}, V({\theta}))}{\tilde{h}({t})}\right].                                                                                                        
\end{eqnarray}
In (\ref{eq:likelihood2}), $\tilde{f}({t}, {y}) = f({y} | {t}) \tilde{h}({t})$,
where $\tilde{h}({t})$ is any density function with support in $\mathbb{R}^{n}$,
and $E_{\tilde{f}}$ denotes
 expectation with respect to $\tilde{f}(\cdot | {y})$. MCML estimates are
then obtained by maximizing
\begin{equation}
\label{eq:mml}
L_{m}({\beta},{\theta}) = \frac{1}{m}\sum_{h=1}^{m}\frac{h({t}_{h}|
                                                          D{\beta}, V({\theta}))}{\tilde{h}({t}_{h})},
\end{equation}
where ${t}_{1},\ldots,{t}_{m}$  are samples from $\tilde{f}(\cdot | {y})$.

The accuracy of the approximation for a given value of $m$ depends critically on
the choice of $\tilde{h}(\cdot)$. A suitable choice is
$h({t}_{h}|D{\beta}_{0}, V({\theta}_{0}))$, where ${\beta}_{0}$ and ${\theta}_{0}$ 
are as close as possible to the maximum likelihood estimates,
$\hat{{\beta}}$ and $\hat{{\theta}}$. In practice, we embed the maximisation of
$L_m({\beta},{\theta})$ within the following
iterative procedure  as suggested in \citet*{geyer1992} and \citet*{geyer1994}:
 let $(\hat{\beta}_1,\hat{\theta}_1)$ denote the values that
maximise $L_m({\beta},{\theta})$ using an initial guess 
at suitable values  $({\beta}_{0},{\theta}_{0})$; repeat the maximisation
with $(\hat{\beta}_1,\hat{\theta}_1)$ replacing $({\beta}_{0},{\theta}_{0})$; continue
until convergence.

For the numerical maximization of \eqref{eq:mml} we use a similar procedure to
the one presented in \citet*{mmle2004}. Write
 $V({\theta}) = \sigma^2 V({\psi})$ where 
$\psi = ( \nu^2/\sigma^2,\tau_1^2/\sigma^2,\tau_2^2/\sigma^2,\phi,\delta,\alpha)^\top$
For a given value of ${\psi}$,
the first and second derivatives
of \eqref{eq:mml} with respect to ${\beta}$ and $\sigma^2$
are analytically tractable
and we use an iterative Newton-Raphson algorithm. We then plug into 
\eqref{eq:mml}
the values $\hat{\beta}(\psi)$ and $\hat{\sigma}(\psi)^2$  and maximize
with respect to ${\psi}$ using direct numerical optimization with a
further re-parameterisation to remove any restrictions on the permissible ranges of 
the parameters; we use a log-transformation
for all elements of $\psi$ except $\alpha$, for which we use
$\log\{(1+\alpha)/(1-\alpha)\}$ to correspond to the
range $-1 < \alpha < 1$. We also consider a variety of starting values
 to guard against false convergence to either a local maximum or an arbitrary point
on a
plateau of the likelihood surface.

We now consider the prediction of 
${T}^* = (T(x_{n+1}),\ldots,T(x_{n+q}))^\top$ at $q$ additional 
prediction locations that are not included in any of the prevalence
surveys. This requires all relevant explanatory variables to
be available at the prediction locations.  We include the mutually
independent random variables $Z_{ij}$ in (\ref{eq:random_part}) 
as part of our target for prediction.  Note that in a linear 
Gaussian geostatistical model, the $Z_{ij}$ would be conflated with Normally 
distributed  measurement errors, whereas in a GLGM for
prevalence survey data the 
analogue of measurement
error is binomial sampling variation and is formally distinguishable
from the extra-binomial variation induced by the $Z_{ij}$.

\citet*{em2002} gives approximate expressions for
 the minimum mean square predictor ${\rm E}[{T}^*|{y}]$ and its variance
 using samples from the conditional distribution
of $T|y$ generated by conditional simulation. For prediction of non-linear functionals of
the prevalence surface,
we first use our MCMC algorithm to generate samples $t_h: h=1,...,m$  from the conditional distribution of $T|y$,
then simulate samples $t_h^*:h=1,...,m$ directly from the multivariate Normal 
conditional distribution of ${T}^* |T= {t}_{h}$.  This has expectation
\begin{equation}
\label{eq:condpredmean}
D^*{\beta} + C^\top V^{-1}({t}_{h}-D{\beta}),
\end{equation}
where $D^*$ is the matrix of covariates at the prediction locations,
and covariance matrix
\begin{equation}
\label{eq:conpredvar}
V^* - C^\top V^{-1} C,
\end{equation}
where
 $V^*$ is the covariance matrix of ${T}^*$ and $C$ is the cross-covariance matrix between ${T}$ and ${T}^*$.
Finally,  we transform the sampled values $t_h^*$ to predicted prevalences,
$$p_h^* = g^{-1}({t}_{h}^*) = (g^{-1}(t_{n+1,h}^*),\ldots,g^{-1}(t_{n+q,h}^*))^\top,$$ 
where $g^{-1}(\cdot)$ is the inverse link function. Typically, the prediction locations will form a fine grid to cover the area of interest, $A$,
so as to approximate
a set of predicted surfaces, ${\cal P}^* = \{p_h^*(x): x \in A\}$ which can then be summarised according to the
needs of each application. For example, we might want to map pointwise means, or selected quantiles,
or predictive probabilities of the exceedance of policy-relevant thresholds.

\section{Simulation study}
\label{sec:simstudy}

We have conducted a simulation study of our proposed methodology
with three aims:  to show that the parameters in (\ref{eq:GLGM}) are identifable;
 to illustrate the
finite sample
properties of the MCML estimators; and to demonstrate the potential gains in predictive performance
that can be obtained by combining data from unbiased and biased surveys.

\subsection{Identifiability and finite sample properties}
\label{subsec:ident}

\begin{figure}
\begin{center}
\vspace*{-0.5in}
\includegraphics[scale=0.6]{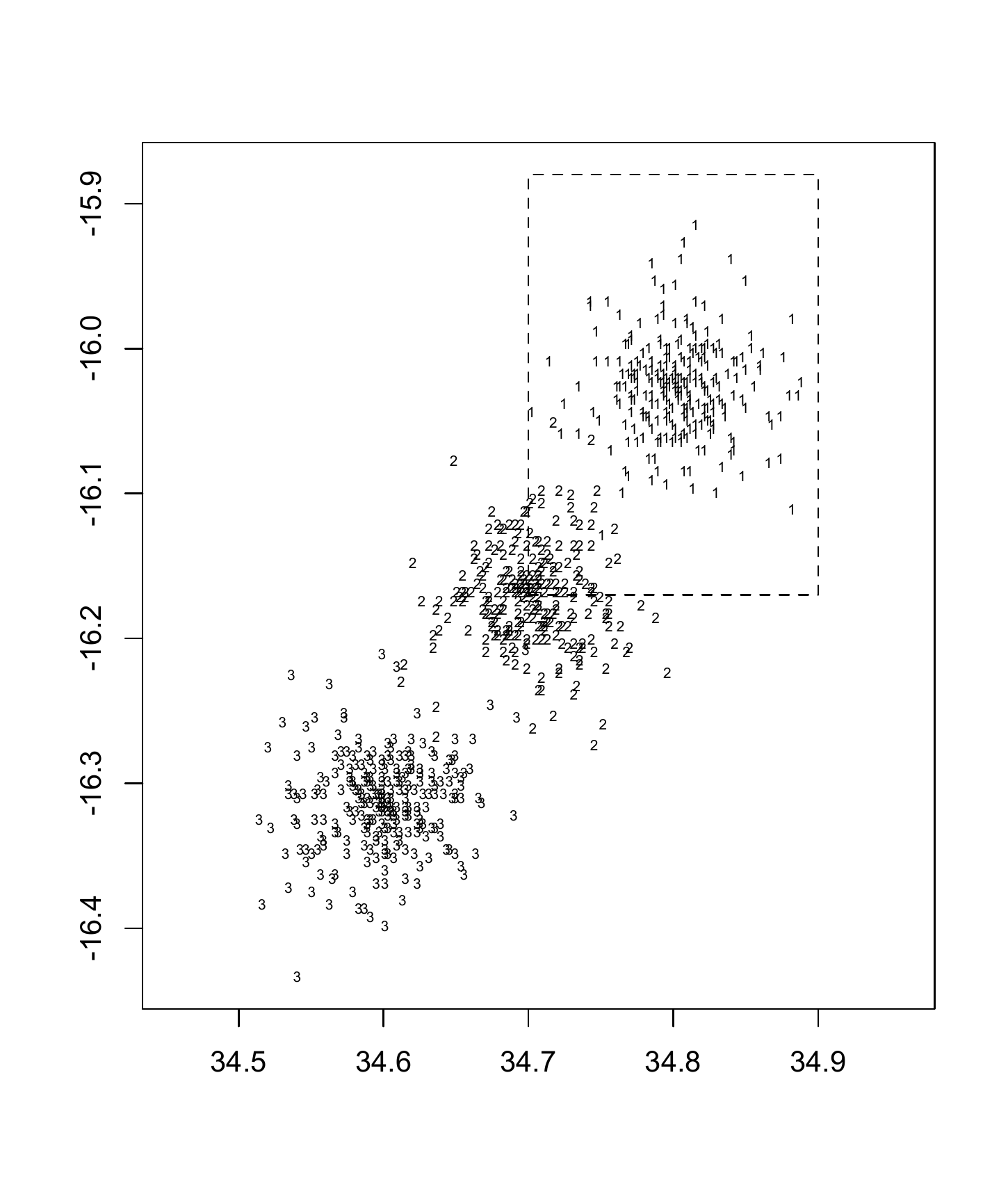}
\caption{example of simulated locations from a biased survey under three different scenarios as defined in Section \ref{subsec:ident}; the dashed lines encompass the region within which locations of an unbiased survey are uniformly generated.\label{fig:ident}}
\end{center}
\end{figure}

For this part of the simulation study we simulated data from two surveys, the first of which was unbiased, the 
second biased. We specified the covariance structure of the model to correspond to the MCML estimates that were obtained in the analysis of malaria prevalence data 
to be reported in Section \ref{sec:application}. We also 
used the same sample sizes as in the malaria application, hence 
 $n_{1} = 425$ (to correspond to the second of the two randomised surveys)
and $n_{2} = 249$ (to correspond to the convenience survey), and the same binomial denominators $m_{ij}$.
We did not use covariates but constant means $\beta_{1}$ for the first survey and $\beta_{1}+\beta_{2}$ for the second survey.
\begin{table}[hbtp]
\centering
\caption{Estimated means and relative biases (RB) of the MCML estimators for the covariance parameters,
and ordered eigenvalues (EV) of their correlation matrix under three scenarios.  \label{tab:ident}}
\vspace{0.6cm}
\begin{tabular}{lr|rrr|rrr|rrr}
   & & \multicolumn{3}{c|}{(1)} & \multicolumn{3}{c|}{(2)} & \multicolumn{3}{c}{(3)} \\
& True value & Mean & RB &  EV & Mean & RB & EV & Mean & RB & EV\\ 
  \hline
$\beta_1$ & 1.000 & 0.997 & -0.003 & 1.677 & 1.011 & 0.011 & 1.677 & 0.998 & -0.002 & 1.811 \\ 
 $\beta_2$ & -1.000 &  -1.011 & 0.011 & 1.287 & -1.013 & 0.013 & 1.425 & -0.980 & -0.020 & 1.472 \\ 
 $\sigma^2$ & 2.186 & 2.132 & -0.025 & 1.173 & 2.093 & -0.042 & 1.298 & 2.005 & -0.083 & 1.141 \\ 
 $\tau^2$ & 0.558 & 0.465 & -0.166 & 0.903 & 0.476 & -0.148 & 0.840 & 0.486 & -0.130 & 0.835 \\ 
  $\nu^2$ & 0.672 & 0.900 & 0.339 & 0.772 & 1.011 & 0.504 & 0.715 & 1.193 & 0.776 & 0.806 \\ 
 $\phi$ & 0.017 & 0.016 & -0.045 & 0.695 & 0.016 & -0.033 & 0.577 & 0.016 & -0.085 & 0.503 \\ 
  $\delta$ & 0.004 & 0.005 & 0.249 & 0.492 & 0.006 & 0.496 & 0.468 & 0.008 & 1.037 & 0.433 \\ 
   \hline
\end{tabular}

\end{table}
We generated the sampling locations for the unbiased survey as an independent random sample from 
the uniform distribution in the rectangle $[34.700, 34.900]\times[-16.170, -15.880]$. The usefulness of 
the data from the biased survey may depend on the degree of overlap between the two sets of sampling locations. For this reason we
generated the sampling locations for the biased survey from each of three inhomogeneous Poisson processes,
 with intensity $\lambda(x) = \exp\{-\|x-x_{0}\|/0.02\}$ and $x_{0}$ set to
each of the three following locations:  $(34.800, -16.025)$,  the center of the the rectangle; 
 $(34.700, -16.170)$, the lower left corner of the rectangle; $(34.600, -16.315)$, a point outside the rectangle. Figure \ref{fig:ident} shows an example of simulated locations for the biased survey under each of these three scenarios.
For each simulation we computed the mean and relative bias of the MCML estimates of the covariance parameters
and the eigenvalues  of their correlation matrix, based on 1000 replications of each of the three scenarios.
The results are shown in Table \ref{tab:ident}.  The estimates
of $\beta_{1}$, $\beta_{2}$, $\sigma^2$, $\tau^2$ and $\phi$ are approximately unbiased
under all three scenarios whereas the estimates of $\nu^2$ and $\delta$,
which relate to the process $B(x)$, become increasingly biased as
the overlap between the two sampled  areas decreases.  
Under all  three scenarios, the 
smallest eigenvalue of the correlation matrix
corresponds to about $6\%$ of its total variation as measured by the sum of
the eigenvalues. Also, the off-diagonal elements of the
 correlation matrix, whose entries are never greater than
 $0.47$ in absolute value, which represents the correlation
between the estimates of $\tau^2$ and $\phi$  in the third scenario.

The overall conclusion from this part of the simulation study is that all
of the model parameters are identifiable, and that the parameter estimates
are approximately unbiased provided that there is a substantial
overlap in the spatial  coverage of the unbiased and biased surveys. This is
as expected,  because without such overlap the two surveys can only
estimate the properties of the sum, $S(x)+B(x)$, in the area covered 
by the biased survey. 

\subsection{Quality variation and temporal variation}
\label{subsec:tqv}

In this part of the simulation study we focus on predictive performance. Our main objective is to indicate to what extent the inclusion of data from
a biased survey can improve predictive inference, under circumstances similar to those that hold in
our malaria application. A secondary objective, as suggested by a reviewer,
is to demonstrate the unreliability of a naive analysis that ignores bias and temporal variation. We therefore 
conducted three analyses of each simulated data-set as follows.
\begin{itemize}
\item Joint (J). The combined data are analysed using the bivariate GLGM as specified in Section \ref{sec:mod_form}.
\item First-survey-only (FSO). Only the data from the first, unbiased survey are used.
\item ``Naive'' (N). The data from the two surveys are analysed using a GLGM that does not account for bias or temporal variation.
\end{itemize}

We consider a {\it quality variation} (QV) scenario, in which one survey is unbiased and
the other biased, and a {\it temporal variation} (TV) scenario, in which both surveys
are unbiased but at different times, with predictions required for the first time period. 

The following features are common to both scenarios. The processes 
$S_i(x):i=1,2$ have mean $\beta_1=1$, 
variance $\sigma^2=1$ and correlation function 
$\rho(u) = \exp(-u/\phi)$ with $\phi=0.15$. 
 Locations of unbiased surveys are uniformly generated in the unit square centred on $x_{0} = (1/2, 1/2)$. Both surveys have the same number of
sampling locations, $n_{1}=n_{2}=300$. 
The binomial denominators
at each sampling location are all set equal to 1.
Our primary focus is on prediction of prevalence at
$x_0$ but we also consider estimation of the parameters 
$\beta_{1}$, $\log \sigma^2$
and  $\log \phi$ that define the model for the underlying prevalence process $S_1(x)$.

In the QV scenario, $S_1(x)=S_2(x)$ for all $x$ and the process $B_2(x)$ has mean $\beta_2=-1$ and correlation function 
$\rho(u) = \exp(-u/\delta)$ with $\delta=0.15$.
Locations from the biased survey are generated from a Poisson process with intensity $\lambda(x) = \exp\{-\|x-x_{0}\|/0.15\}$ so that points closer to $x_{0}$ are more likely to be sampled, as might occur when using a convenience sampling strategy and
$x_0$ is the location of a health-care facility. Finally,  
we consider four values, $\nu^2=0.5, 1, 2, 4 $, for the variance of the
process $B_2(x)$,  corresponding to increasingly severe 
 spatial variation in the bias.

In the TV scenario, the cross-correlation function between
$S_1(x)$ and $S_2(x)$ is $\alpha \exp(-u/\phi)$. We consider 
three values, $\alpha=0.2,0.5,0.8$, to correspond to weak, moderate
and strong correlation between the two prevalence surfaces.

 The results are summarised in Tables \ref{tab:qv} and \ref{tab:tv}.  These show
 estimates of the root-mean-square-error (RMSE)
 and coverage of nominal $95\%$ confidence intervals
(CIC) for MCML estimates of the parameters $\beta_{1}$, $\log \sigma^2$
and  $\log \phi$,
and for the minimum mean square error predictors of 
$S_{1}(x_{0})$ and $\beta_{1}+S_{1}(x_{0})$. Each
entry is calculated from 1000 independent replicates of the simulation
model. 

\begin{table}[htp]
\caption{Estimated RMSE, bias, SD and $95\%$ CIC for the MCML estimates of $\beta_{1}$, $\log\sigma^2$, $\log\phi$, for the minimum mean square error predictor of $S_{1}(x_{0})$ at location $x_{0}$ and $\beta_{1}+S(x_{0})$, under QV scenarios. \label{tab:qv}}
\centering
\small
\vspace{0.6cm}
\begin{tabular}{rcrrrr|rrrr}
Model & Parameter & \multicolumn{4}{c}{RMSE} &  \multicolumn{4}{c}{CIC}\\
   & & \multicolumn{4}{c|}{$\nu^2$} &\multicolumn{4}{c}{$\nu^2$} \\
& & 0.5 & 1 & 2 & 4 & 0.5 & 1 & 2 & 4 \\ 
  \hline
J & & 0.37 & 0.36 & 0.36 & 0.36 & 0.95 & 0.95 & 0.95 & 0.94 \\ 
  FSO & $\beta_{1}$& 0.37 & 0.36 & 0.35 & 0.35 & 0.95 & 0.95 & 0.95 & 0.95 \\ 
  N & & 0.50 & 0.52 & 0.51 & 0.48 & 0.82 & 0.79 & 0.83 & 0.77 \\ 
  \hline
  J & & 0.94 & 0.60 & 0.63 & 1.33 & 0.99 & 0.94 & 0.95 & 0.99 \\ 
  FSO & $\log\sigma^2$& 1.08 & 1.14 & 1.04 & 0.99 & 0.97 & 0.97 & 0.97 & 0.97 \\ 
N & & 0.78 & 0.49 & 0.52 & 0.54 & 0.99 & 0.96 & 0.95 & 0.91 \\ 
  \hline
  J & & 0.84 & 0.81 & 0.98 & 0.95 & 0.92 & 0.90 & 0.92 & 0.92 \\ 
 FSO & $\log\phi$& 1.45 & 1.42 & 1.32 & 1.44 & 0.94 & 0.94 & 0.92 & 0.94 \\ 
  N & & 0.69 & 0.68 & 0.66 & 0.70 & 0.92 & 0.90 & 0.88 & 0.83 \\ 
  \hline
 J & & 0.80 & 0.79 & 0.88 & 0.86 & 0.95 & 0.95 & 0.95 & 0.95 \\ 
  FSO & $S_{1}(x_{0})$& 0.91 & 0.86 & 0.92 & 0.85 & 0.95 & 0.95 & 0.95 & 0.94 \\ 
  N & & 0.93 & 0.96 & 1.13 & 1.21 & 0.80 & 0.80 & 0.77 & 0.76 \\ 
  \hline
    J & & 0.72 & 0.73 & 0.82 & 0.83 & 0.95 & 0.95 & 0.95 & 0.95 \\ 
  FSO & $\beta_{1}+S_{1}(x_{0})$ & 0.83 & 0.80 & 0.84 & 0.81 & 0.95 & 0.94 & 0.95 & 0.94 \\ 
  N & & 1.10 & 1.15 & 1.33 & 1.35 & 0.78 & 0.78 & 0.75 & 0.74 \\ 
   \hline
\end{tabular}
\end{table}
\begin{table}[htp]
\caption{Estimated RMSE, bias, SD and $95\%$ CIC for the MCML estimates of $\beta_{1}$, $\log\sigma^2$, $\log\phi$, for the minimum mean square error predictor of $S_{1}(x_{0})$ at location $x_{0}$ and $\beta_{1}+S(x_{0})$, under TV scenarios. \label{tab:tv}} 
\centering
\small
\vspace{0.6cm}
\begin{tabular}{rcrrr|rrr}
Model & Parameter & \multicolumn{3}{c}{RMSE} & \multicolumn{3}{c}{CIC} \\
   & & \multicolumn{3}{c|}{$\alpha$} & \multicolumn{3}{c}{$\alpha$} \\
   & & 0.2 & 0.5 & 0.8 & 0.2 & 0.5 & 0.8 \\
  \hline
J & & 0.35 & 0.35 & 0.35 & 0.95 & 0.93 & 0.94 \\ 
 FSO & $\beta_{1}$& 0.36 & 0.36 & 0.35 & 0.94 & 0.93 & 0.93 \\ 
  N & & 0.63 & 0.63 & 0.64 & 0.29 & 0.38 & 0.42 \\ 
  \hline
  J & & 0.60 & 0.68 & 0.69 & 0.94 & 0.94 & 0.93 \\ 
  FSO & $\log\sigma^2$& 1.04 & 1.46 & 1.47 & 0.97 & 0.96 & 0.96 \\ 
  N & & 1.18 & 0.95 & 0.87 & 0.91 & 0.92 & 0.95 \\ 
  \hline
  J & & 0.93 & 0.92 & 0.91 & 0.91 & 0.93 & 0.93 \\ 
  FSO & $\log\phi$ & 1.47 & 1.56 & 1.75 & 0.92 & 0.93 & 0.94 \\ 
  N & & 1.32 & 1.09 & 1.02 & 0.92 & 0.94 & 0.94 \\ 
  \hline
  J & & 1.37 & 1.30 & 1.28 & 0.95 & 0.95 & 0.94 \\ 
  FSO & $S_{1}(x_{0})$& 1.37 & 1.30 & 1.31 & 0.94 & 0.93 & 0.92 \\ 
  N & & 1.39 & 1.32 & 1.28 & 0.87 & 0.91 & 0.92 \\ 
  \hline
    J & & 1.34 & 1.28 & 1.26 & 0.96 & 0.96 & 0.94 \\ 
  FSO & $\beta_{1}+S_{1}(x_{0})$ & 1.35 & 1.27 & 1.27  & 0.95 & 0.95 & 0.94 \\ 
  N & & 1.50 & 1.39 & 1.36 & 0.86 & 0.90 & 0.91 \\ 
   \hline
\end{tabular}
\end{table}

Overall, J outperforms FSO, which in turn outperforms N. Under the QV scenario, the main benefits of J are in the prediction of $S_{1}(x_{0})$ for values of $\nu^2$ smaller than $4$. The N approach yields much higher values of RMSE for the estimates of $\beta_{1}$, $S_{1}(x_{0})$ and $\beta_{1}+S_{1}(x_{0})$ and
very poor CIC.
 Under the TV scenario, the biggest gains achieved by J over FSO are in estimating
 the parameters $\log \sigma^2$ and $\log \phi$. Both J and FSO
perform similarly with respect to prediction of $S_{1}(x_{0})$ and $\beta_{1}+S_{1}(x_{0})$. The N approach, which in this scenario
consists of combining the data
under the assumption that  $S_{1}(x)=S_{2}(x)$ for all $x$,
i.e. $\alpha=1$, has the worst performance.

\section{Application: malaria prevalence mapping}
\label{sec:application}

 In this Section, we
use our proposed methodology to construct malaria prevalence maps
for an area of Malawi by combining information from three surveys.
All three surveys
were directed at the same target population, covering a 400 square km area 
within Chikhwawa District, Southern Malawi. 
 Two of the surveys were ``rolling'' Malaria
Indicator Surveys (MIS) \citep*{roca2012}, 
that used two different  practical strategies to obtain random, and 
therefore unbiased, samples
from the population at risk.
The third was a facility-based  survey  that used
 a convenience sampling strategy, in which
 recruitment took place at a central child-vaccination
clinic at the main hospital in the centre of the study area.
We refer to this as the Easy Access Group (EAG)
study.

\subsection{Data}
\label{subsec:data}

\begin{figure}
\begin{center}
\includegraphics[scale=0.52]{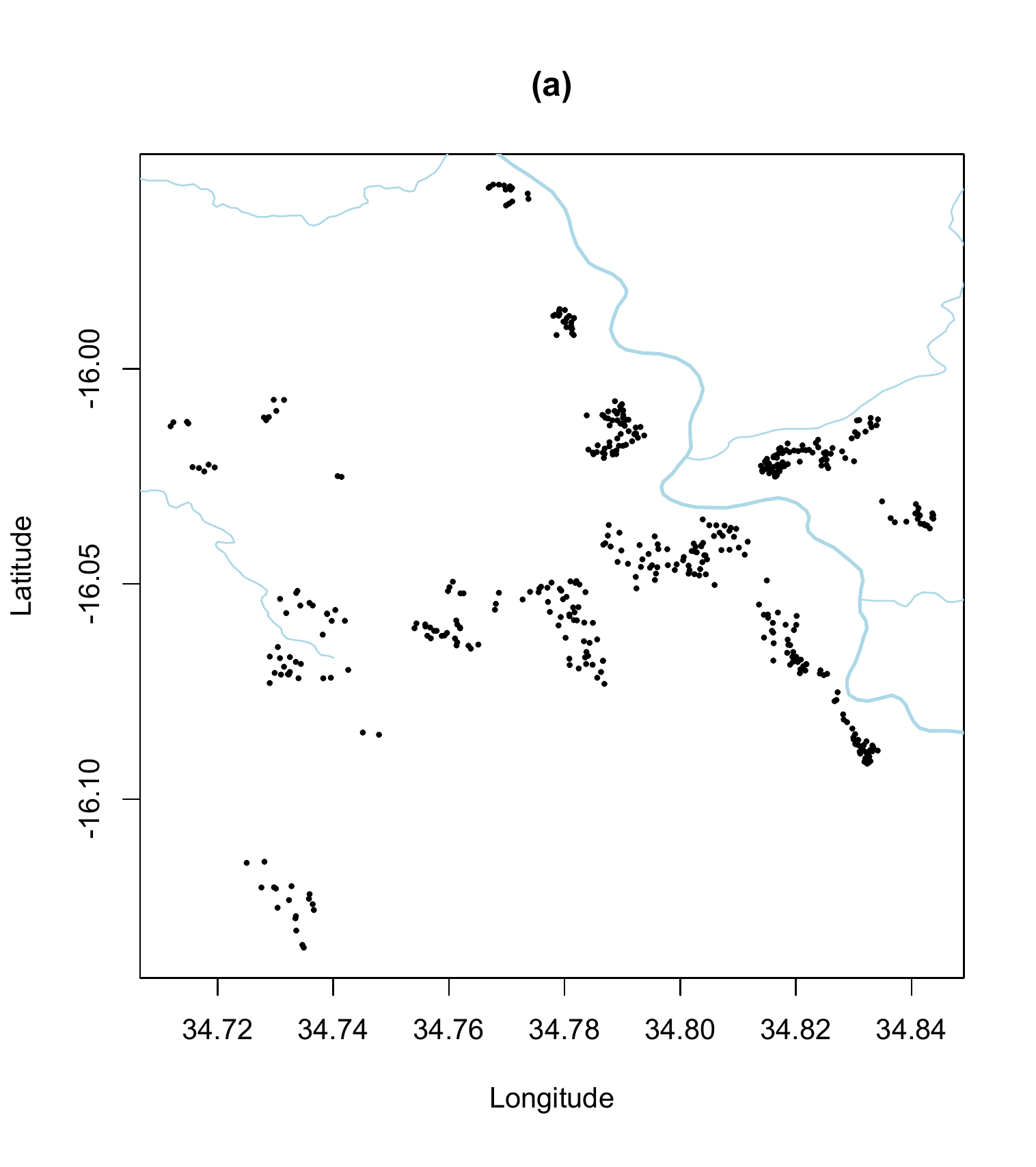}
\includegraphics[scale=0.52]{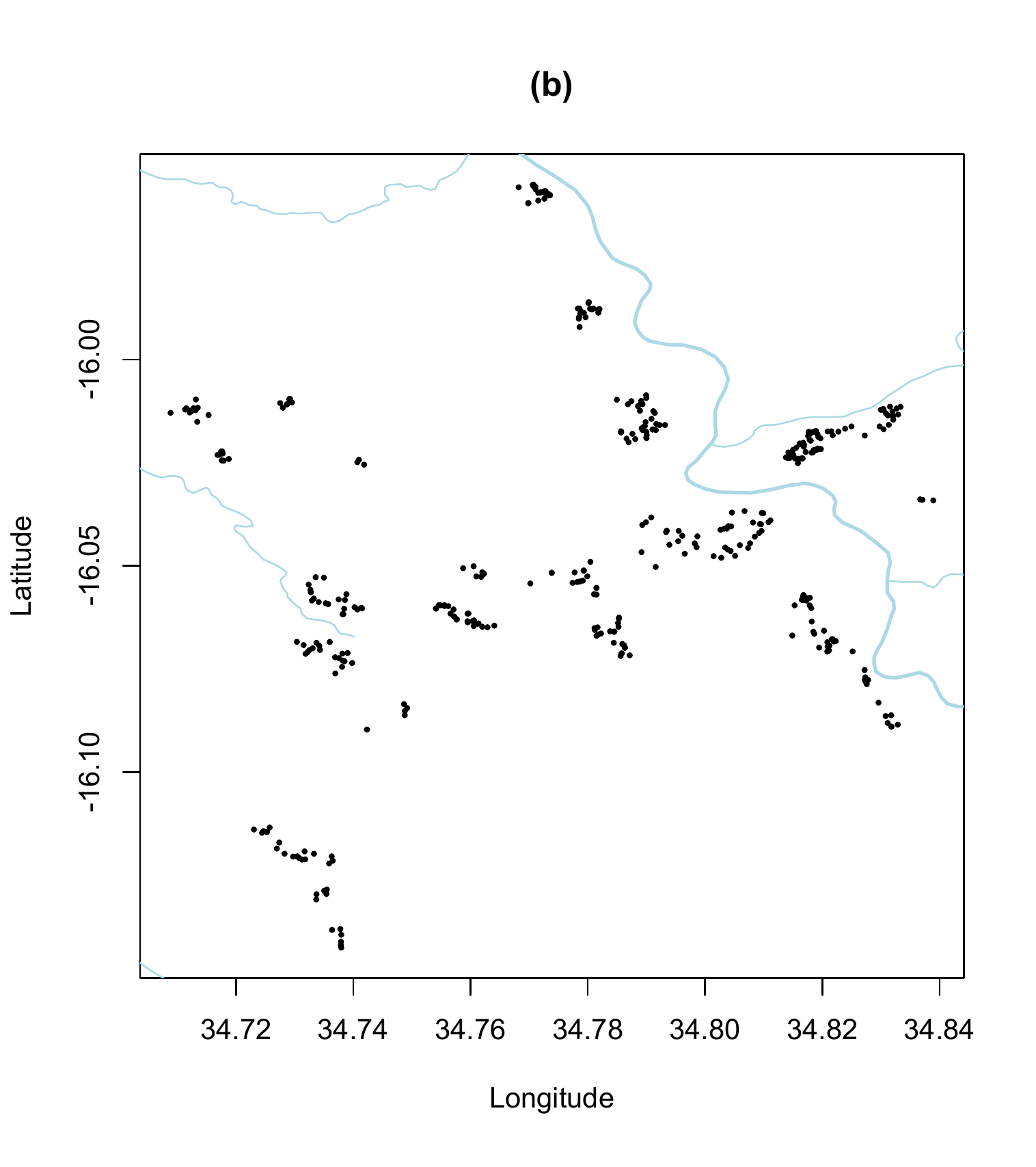}
\includegraphics[scale=0.52]{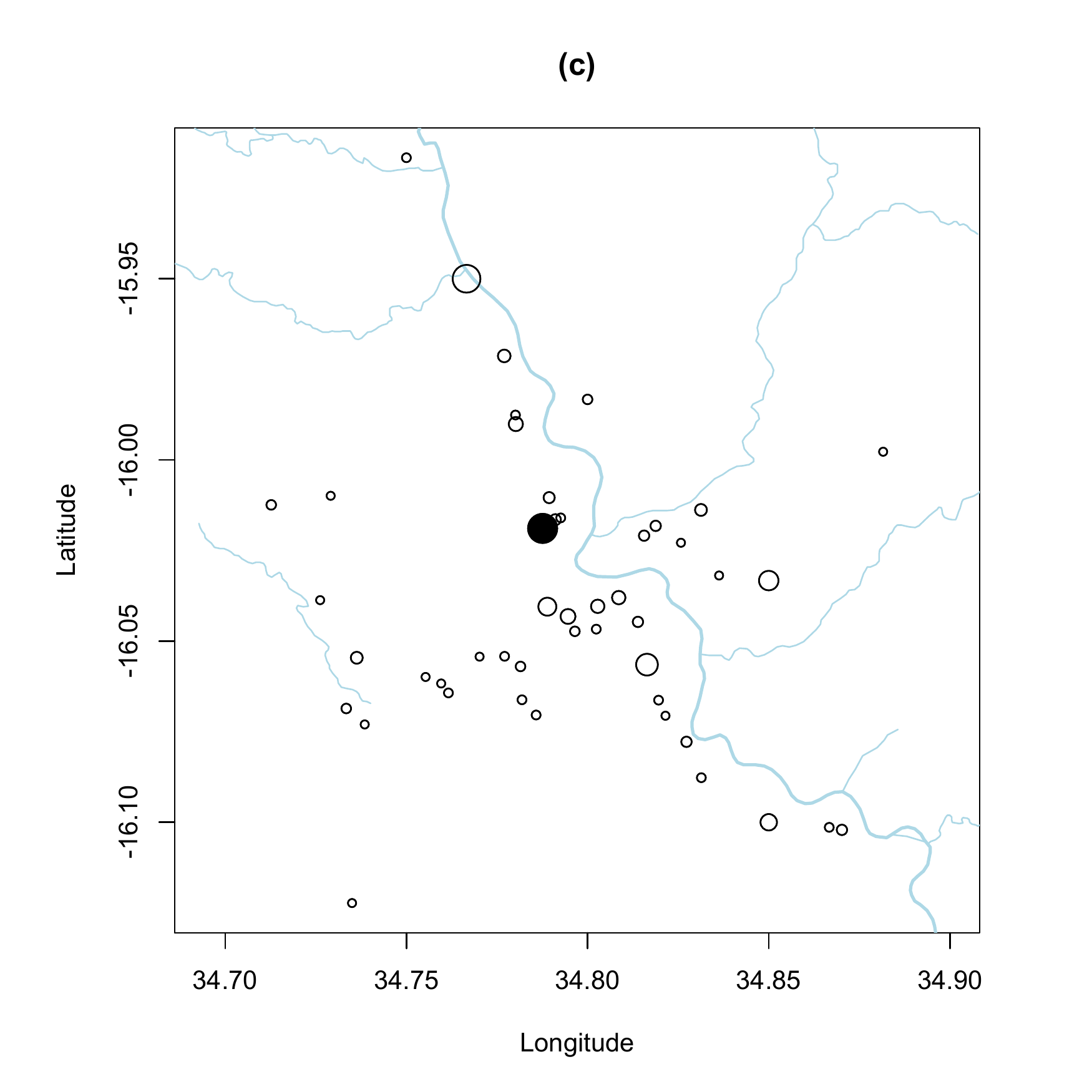}
\caption{Sampled locations for (a) rMIS, (b) eMIS and (c) EAG. 
Coordinates are of individual houses  for rMIS and eMIS,
and of the villages for EAG; in (c), the radius of each circle is proportional to the number of the sampled households from the respective village and the black solid point corresponds to CDH village. The light blue lines represent waterways, with the thicker line corresponding to Shire river. \label{fig:coords}}
\end{center}
\end{figure}
Two population-level continuous malaria indicator surveys were conducted over the
period May
2010 to April 2012. Both surveys recruited
children aged less than five years in a sample of 50 village communities in order to 
monitor the malaria intervention coverage and childhood burden of malaria in 
a designated area 
containing the sampled villages, which was chosen to represent the catchment area of
the Chikhwawa District Hospital (CDH). The two surveys differed in
the  sampling strategy used, as described below. We
refer to these two surveys  as the rMIS, covering the period May 2010 to April
2011 and the eMIS, covering the period May 2011 to April 2012. Throughout the two-year period
seven or eight villages were randomly selected
per month so as to sample all 50 villages twice yearly, once during the high-transmission season and once
during the low-transmission season. Within sampled villages, selection of households 
was as follows. In the
rMIS, households were randomly selected within each village from a list of households, with
sampling probability proportional to village population size, based on a population enumeration
exercise. In the eMIS, a more economical
 ``spin-the-bottle'' method was used to identify a random set of households within villages. A
bottle was placed in the center of a village and used
to select random directions. A virtual line was drawn in each chosen direction to the border of the
village, the households that intersected this line were counted, and from 
these a random household number was chosen as the starting point. The number of houses selected within each village was proportional to the estimated village population size. 
Figures \ref{fig:coords} (a)-(b) show the sampled locations for the rMIS and the eMIS. 

The third survey is a continuous facility-based MIS in children attending the immunization clinic at the CDH, conducted from May 2011 to April 2012. The objective of this study was to determine if estimates of uptake of control interventions and the burden of malaria from convenience sampling 
were comparable to those from a randomised MIS conducted
 within the same catchment area of CDH. Children from 3 months of age who attended the vaccination clinic, and any accompanying sibling below 5 years, were recruited. 
Between  30 and 50 children were recruited per month. Village of origin was extracted by direct questioning. If the village was not one of the 50 eMIS/rMIS villages for which the location was already known, its coordinates were determined retrospectively. The results for villages within 15km of CDH were extracted to make the 
catchment area of the EAG comparable to that of the rMIS and eMIS.
Malaria control efforts by the national control program during the first period included a
district-wide household indoor residual spraying campaign between February and April 2011.
Practical difficulties resulted in this campaign 
being conducted at the end of the rainy season rather than, as would have been 
ideal, before the
start of the rainy season. This will have reduced its potential impact. Insecticide-treated net
control efforts were stable over the three months of the campaign, with distribution to women attending antenatal clinics
and mother and child clinics. 

\subsection{Results}
\label{subsec:results}

The response from each child was a binary indicator of the outcome of 
a rapid diagnostic test (RDT) used to test for the presence of malaria from a finger-prick blood sample.
Six
explanatory variables were considered, as
defined in Table \ref{tab:explanatory_variables}. Socio-Economic-Status (SES), an indicator of household wealth taking discrete values from 1 (poor) to 5 (wealthy),
was derived by an application of principal component analysis as discussed in \citet*{vyas2006}.

\begin{table}[htp]
\caption{Explanatory variables used in the analysis of the Chikhwawa malaria prevalence surveys}
\begin{center}
\begin{tabular}{ll} \hline
1 & intercept \\
2 & at least one treated bed-net in the household (yes/no)\\
3 & indoor residual spraying in the past two months (yes/no) \\
4 & high-transmission season (January-June/July-December) \\
5 & distance from the closest waterway (km) \\
6 & Socio-Economic-Status (SES, 1 to 5)\\ \hline
\end{tabular}
\label{tab:explanatory_variables}
\end{center}
\end{table}

It was thought  that health facility utilization might be  associated with SES
as previously observed in \cite{gahutu2011}, where children with relatively
high SES were  more likely to attend 
a CDH.  Table \ref{tab:SES} shows the average SES observed in each
of our  three surveys. Enrolled children in 
the EAG study show a higher average SES then those in the two other surveys. Additionally, Table \ref{tab:SESandRDT} shows that the relationship
between SES and the distribution of the number of  RDT positive results 
per household differs 
 between the two gold-standard surveys and the convenience survey.
We therefore allowed SES to have a direct effect on the spatially 
structured bias  of the EAG survey in addition to its possible association with prevalence.
\begin{table}[ht]
\centering
\tabcolsep = 0.1cm
\caption{Mean and standard deviation (SD) of SES in the three surveys.\label{tab:SES}}
\vspace{0.5cm}
\begin{tabular}{rrrr}
  \hline
  & \multicolumn{3}{c}{SES}\\
 & rMIS & eMIS & EAG\\ 
  \hline
Mean & 2.76 & 2.50 & 3.45\\ 
  SD & 1.45 & 1.37 & 1.39 \\ 
   \hline
\end{tabular}
\vspace{1cm}
\caption{Distribution (percentage) of the number of positive RDTs 
per household for each value of  SES, in the
convenience survey (EAG, left-columns) and in 
the gold-standard surveys (rMIS and eMIS, right columns)
 \label{tab:SESandRDT}}
\vspace{0.5cm}
\begin{tabular}{crrrrrr|rrrrr}
 & & \multicolumn{5}{c}{SES (EAG)} & \multicolumn{5}{c}{SES (rMIS and eMIS)} \\
 &  & 1 & 2 & 3 & 4 & 5 & 1 & 2 & 3 & 4 & 5 \\ 
  \hline
  & 0 &75.76 & 80.56 & 77.50 & 79.10 & 89.04 & 54.58 & 63.09 & 71.72 & 73.29 & 83.74\\ 
 RDT & 1 & 21.21 & 19.44 & 20.00 & 20.90 & 10.96 & 40.49 & 33.56 & 25.25 & 22.60 & 15.45 \\ 
  positives& 2 & 3.03 & 0.00 & 2.50 & 0.00 & 0.00 & 4.58 & 3.35 & 3.03 & 3.42 & 0.81\\ 
  & 3 & 0.00 & 0.00 & 0.00 & 0.00 & 0.00 & 0.35 & 0.00 & 0.00 & 0.69 & 0.00  \\ 
  \hline
\end{tabular}
\end{table}

\begin{table}[ht]
\caption{Monte Carlo maximum likelihood estimates and $95\%$ confidence intervals. \label{tab:estim}}
\begin{center}
\begin{tabular}{lrr}
  \hline
 Term & Estimate & 95\% confidence interval \\ 
  \hline
$\beta_{1}$ & -0.272 & (-1.382,  0.862) \\ 
$\beta_{2}$ & -0.439 & (-0.623, -0.277) \\ 
$\beta_{3}$ & -0.399 & (-0.621, -0.189)\\ 
$\beta_{4}$ & 0.415 & (0.206, 0.598)\\ 
$\beta_{5}$ & -0.373 & (-0.970,  0.116) \\ 
$\beta_{6}$ & -0.151 & (-0.233, -0.072)\\ 
$\beta_{7}$ & -0.096 & (-0.222,  0.021) \\ 
$\sigma^2$ & 2.186 & (0.955,  3.155)\\ 
$\tau^2$ & 0.558 & (0.089,  1.231) \\ 
$\nu^2$ & 0.672 & (0.525,  0.802)\\ 
$\alpha$ & 0.859 & (0.483,  0.924)\\ 
$\phi$ & 0.017 & (0.006,  0.032) \\ 
$\delta$ & 0.004 & (0.001,  0.025)\\ 
   \hline
\end{tabular}
\end{center}
\end{table}

The resulting model for the combined data therefore included
seven regression parameters, $\beta_1, \beta_2, ... , \beta_7$. Let
 $\beta^\top = (\beta_{1}, \dots, \beta_{6})$ and denote
by  $d(x_{ij})$  the vector of covariates associated with location $x_{ij}$.
Use $i=1,2,3$ to denote
rMIS, eMIS and EAG, respectively. Then, the linear predictor is
\begin{eqnarray*}
\eta_{ij} &=& d(x_{ij})^\top\beta + S_{i}(x_{ij}) + I(i = 3)[B(x_{ij})+\beta_{7}\text{SES}_{ij}]+Z_{ij}, \\\
&& i=1,\ldots, 3; j = 1,\ldots, n_{i},
\end{eqnarray*}
where $n_{1} = 475$, $n_{2}=425$ and $n_{3}=249$. 
Note that in the joint model for $S_1(x)$, $S_2(x)$ and
$S_3(x)$, $\alpha_{23} = 1$ because the EAG study took place 
over the same period as the eMIS.
We therefore use $\alpha$ to denote $\alpha_{12}$ and set $S_{3}(x) = S_{2}(x)$. 
We also assume equal variances $\tau^2$
 for the nugget term $Z$ across all three
surveys. Finally, we define $\cov\{S_{1}(x), S_{2}(x')\} = \sigma^2\alpha\exp\{-\|x-x'\|/\phi)$ where $\sigma^2 > 0$, $\phi > 0$ and $-1 < \alpha < 1$. \par
\begin{figure}[h]
\begin{center}
\includegraphics[scale=0.48]{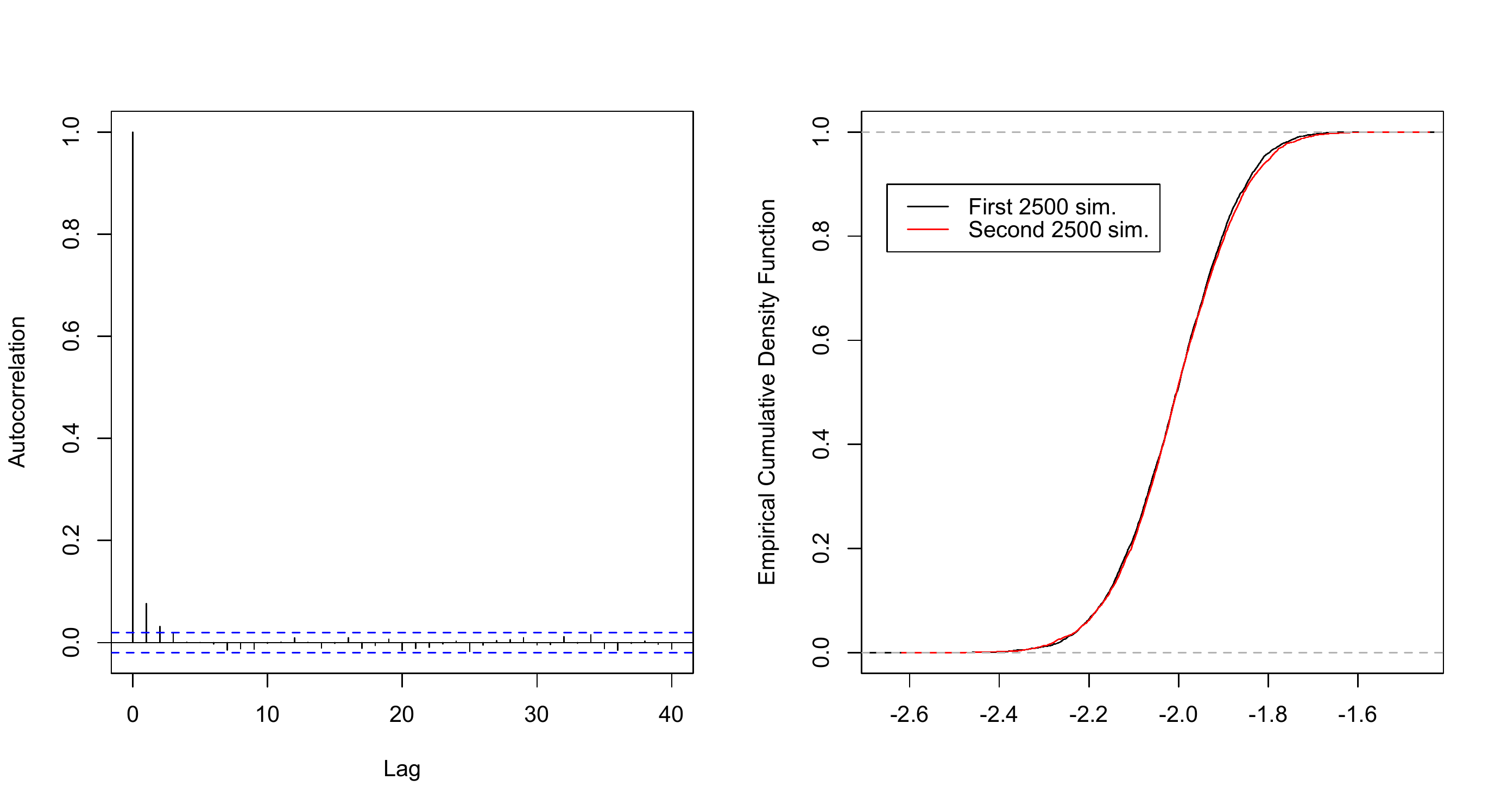}
\caption{Diagnostic plots for the convergence of the posterior average of the random effect. Left panel: correlogram of the 5000 simulated values. Right panel: empirical cumulative density function for the first (black line) and second (red line) 2500 simulated values. \label{fig:diag}}
\end{center}
\end{figure}

Table \ref{tab:estim} shows the Monte Carlo maximum likelihood estimates 
of the model parameters
together with $95\%$ confidence intervals. Each evaluation of the log-likelihood used $5000$ simulated values, obtained by conditional simulation of $110000$ values
and sampling every $20$th realization after
discarding a burn-in of $10000$ values. Figure \ref{fig:diag} shows two diagnostic plots for the average random effect: convergence of the MCMC algorithm appears to be 
satisfactory.

The confidence intervals in Table \ref{tab:estim} were 
calculated using the following parametric bootstrap procedure.
Using the parameter estimates in Table \ref{tab:estim} we simulated $1000$ data-sets
 from the model,
applied to each simulated data set the Monte Carlo maximum likelihood method
with $5000$ conditional simulations, and 
 computed the empirical quantiles of the $1000$ 
resulting estimates of each parameter. 
Although this procedure introduces additional Monte Carlo error, it allows us to compute confidence intervals without relying on 
questionable Normal approximations for the distribution of 
the Monte Carlo maximum likelihood estimates. \par
From Table \ref{tab:estim}, we see
 that the ownership of at least one treated bed net, the presence of residual indoor spraying and an increase in SES are all associated with a reduction in the prevalence of a positive RDT. The distance from the closest waterway is not significant, although the sign of the regression coefficient suggests
 that prevalence  decreases with increasing distance. The period January to June,
which is known to be a period of high malaria transmission, 
 is associated with a significant increase in prevalence, by an estimated factor of
$\exp(0.415) \approx 1.51$.

\begin{figure}[h]
\begin{center}
\includegraphics[scale=0.91]{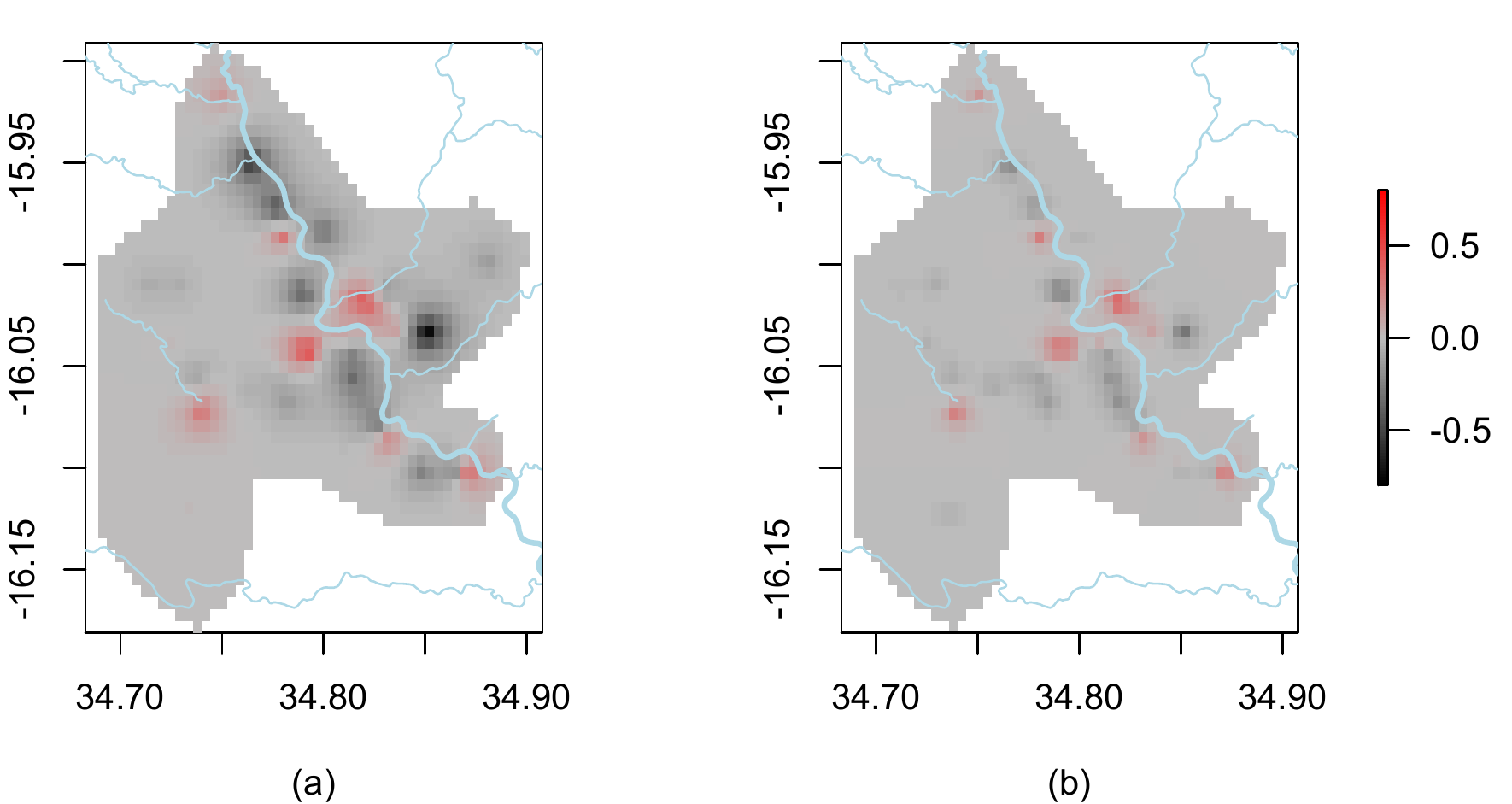}
\end{center}
\caption{Predicted bias surface $B(x)$ (a) without the interaction term of SES and (b) including the effect of SES on the spatial bias. \label{fig:bias}}
\end{figure}

The regression coefficient $\beta_7$, which represents the
additional effect of SES on the bias of the EAG data,
is not significant, but its inclusion nevertheless
makes a noticeable difference to the
 predicted bias surface.  Figures \ref{fig:bias}(a) and \ref{fig:bias}(b) show
the minimum mean square error predictions of the bias with and without
including the regression on SES.

\begin{figure}[h]
\centerline{\includegraphics[scale=0.7]{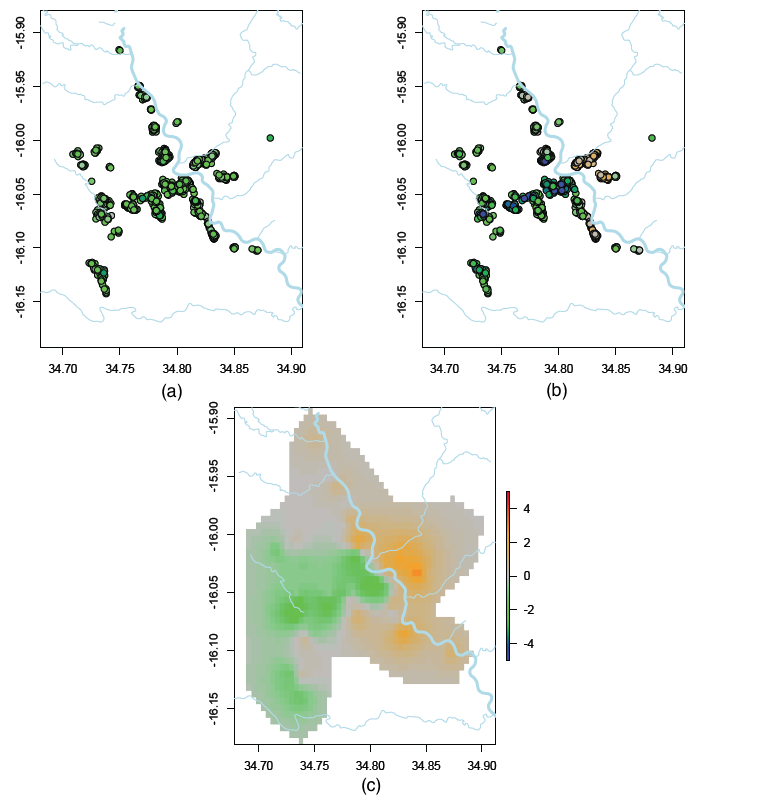}}
\caption{Predictions of (a) $d(x_{2j})^\top\beta$ and (b) $d(x_{2j})^\top\beta + S_{2}(x_{2j})$ at observed locations; (c) predicted surface of the unexplained spatial variation $S_{2}(x)$. The same colour scale has been used for the point predictions.\label{fig:pred}}
\end{figure}

The estimate $\hat{\alpha}=0.859$, albeit with a wide confidence interval,
 indicates a 
strong correlation between prevalences in the two sampling
periods, 2010-2011 and 2011-2012. 

 Figures \ref{fig:pred}a and b show
 the contributions of the linear regression and of the unexplained spatial variation
to the predicted  log-odds of prevalence at each of the observed locations. Figure \ref{fig:pred}c shows the unexplained component, $\hat{S}(x)$,
of the predicted prevalence as a spatially continuous surface.  The clear and
substantial difference between adjacent areas to the east and west of the river Shire strongly suggests the existence of one, or
more, social or environmental risk-factors that are not captured
by the available explanatory variables. 

\begin{figure}[h]
\begin{center}
\includegraphics[scale=0.8]{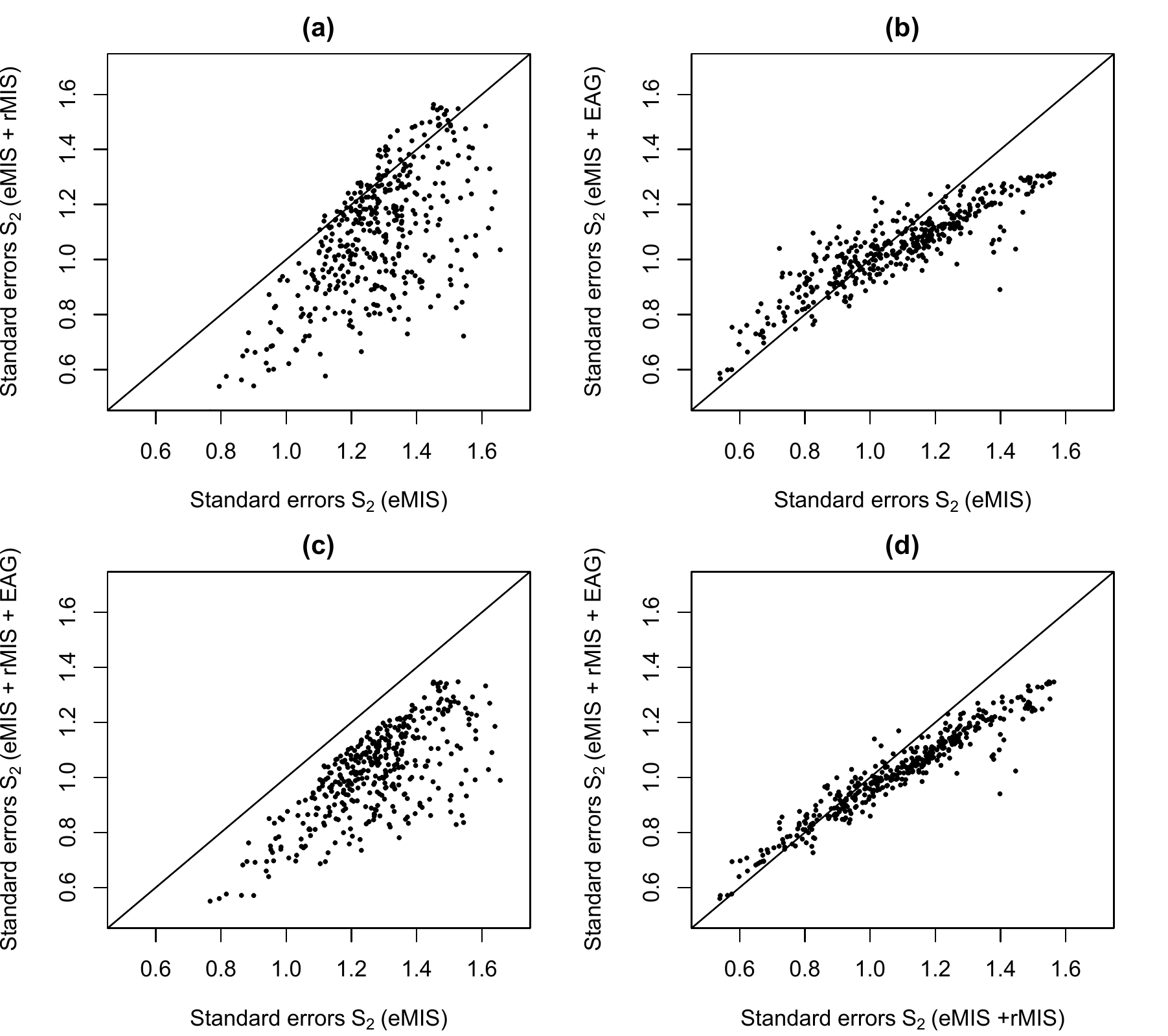}
\end{center}
\caption{Scatter plots of the prediction standard errors for $S_{2}(x)$ 
at sampled locations $x$, using models fitted to the data from: (a) eMIS against eMIS and rMIS; (b) eMIS against eMIS and EAG; (c) eMIS, rMIS and EAG against eMIS; (d) eMIS, rMIS and EAG against eMIS and rMIS.\label{fig:std_errors}. The solid line represents the identity line.}
\end{figure}

Figure \ref{fig:std_errors} shows pairwise scatter plots to compare the
prediction standard deviations for $S_{2}(x)$ at the sampling locations.
Figure \ref{fig:std_errors} (a) shows that analysing rMIS and eMIS data in the joint model 
for temporal variation results in substantially better precision than using only the eMIS;  
Figures \ref{fig:std_errors} (c) and (d) 
show the further, but more modest, gains resulting from addition of the 
data from the EAG; in contrast, Figure \ref{fig:std_errors} (b) suggests little or no benefit
from adding the EAG data to the eMIS data, with predictive standard deviations decreasing at
some locations but increasing at others. 

\section{Discussion}
\label{sec:discussion}

We have developed a class of multivariate GLGMs for the combined data from 
multiple spatially referenced surveys, and 
associated Monte Carlo methods for
 maximum likelihood estimation and spatial prediction within the
proposed class of models. 

The model as defined by (\ref{eq:GLGM}) is
the minimally parameterised model that captures the essential features of our motivating
application: variation in 
data-quality arising from non-randomised sampling;
variation in prevalence over time; binomial and extra-binomial
sampling variation.  We have shown that 
all of the
model parameters are identifiable from surveys of comparable size to the ones available
to us for the application. If substantially
larger data-sets were available, it would be of interest to extend the model in various
ways, for example by relaxing the assumption of common parameters for
 the prevalence surfaces $S_i(x)$ at different times or by allowing cross-correlation
between the $S_i(x)$ and their paired bias surfaces $B_i(x)$.
Additionally, if a large number of surveys were conducted at irregularly spaced
time-points within partly overlapping time periods, the use of a structured spatio-temporally
continuous process $S(x,t)$, as mentioned in Section \ref{sec:mod_form}, would
be more appealing than a discrete set of processes $S_i(x)$ at specific times $t_i$.

The Monte Carlo maximum likelihood
estimation procedure is computationally intensive, primarily because of the need to 
use parametric bootstrapping to compute standard errors reliably. For this reason, we are currently developing a much faster
Monte Carlo method for approximate evaluation of the likelihood function.

In our application to malaria prevalence surveys, we 
combined data from three surveys, two of which 
were unbiased and conducted in two consecutive years,whilst the third
was a potentially biased convenience survey conducted over the same time-period as the
second unbiased survey. We obtained substantial gains in the precision of
spatial predictions by combining the data from the two unbiased surveys
and further, but smaller, gains from combining the data from all three surveys.

One of the limitations of our approach is that it assumes that 
at least one of the available surveys represents an unbiased gold-standard. This is a
reasonable assumption when, as in our application, at least one of the surveys uses a 
properly randomised sampling scheme. When we cannot assume that one of the surveys
is unbiased by design, it is difficult to see how any method could deliver reliable
predictions without additional assumptions that would be difficult or impossible
to validate
empirically. 

The problem that we have addressed in this paper is related to, but distinct from,
the problem of preferential sampling as formulated in \cite{prefer2010}. In both
settings, the goal is to predict the realisation of a latent spatial process $S(x)$
using data obtained by a potentially biased sampling scheme.  In preferential
sampling, the bias arises from a direct relationship between the value of $S(x)$ and
the probability that the location $x$ will be sampled. In the present paper, the
bias is a function of the location $x$ itself, rather than of the value of $S(x)$. In
the context of disease prevalence mapping, a further distinction is between
properties of a location and properties of a person who happens to live at that location.  Thus,
in our application a relationship between a child's location and the
likelihood that they would present at the CDH would not, in itself, result in bias. Rather, 
the bias surface $B(x)$ allows for the possibility that the sub-population
of children who present at the CDH differs from the general population 
with respect to their exposure to unmeasured risk-factors
for malaria.

Our approach is of potentially wide application to disease monitoring and control
in low-resource settings, where registry data are typically not available.
 The ability to combine data from surveys that vary in their level of bias and 
timing can inform more accurate, local-area burden maps,
allowing for improved risk stratification of high burden areas and identification of 
transmission hot-spots. For example, although substantial progress has been made over the past decade with malaria control by homogeneous scaling up of interventions at national level, it is increasingly recognized by funders and policy makers that a more targeted approach focused on high-burden areas or hot-spots may be more  
cost-effective. Furthermore, apart from its potential to optimize the use of available data, 
our approach can also inform improved prospective data collection, 
by using the fitted model 
in simulation studies to identify
efficient prospective hybrid sampling approaches 
that combine convenience and random sampling strategies
in ways that acknowledge and exploit spatial and/or temporal heterogeneity as 
revealed by analyses of the kind described in Section \ref{sec:application}.

In conclusion, our proposed approach provides a way of making use of
mixed source prevalence data to improve estimates of spatial predictions.
These are urgently needed to
support control programmes and develop more accurate local spatio-temporal risk stratification maps that can
inform more targeted control efforts. Malaria is one of a number of diseases
that bring a high public health burden in 
low-resource settings, whilst exhibiting
 highly heterogeneous distributions
 across space and time. Control of such diseases needs methods of
 continuous monitoring of prevalence
and evaluation of control measures that make  the best possible use of
limited resources, and 
will  therefore benefit greatly from
the ability to combine national household surveys with more local convenience sampling strategies
without compromising the validity of the resulting prevalence estimates.

\section*{Acknowledgements}
 We thank the participants and staff involved in the data collection of the presented household prevalence surveys, especially Dr Roca-Feltrer who oversaw the survey field teams between 2010 and 2011  

\section*{Funding}
Emanuele Giorgi holds an ESRC-NWDTC funded Ph.D. studentship. Dr Sanie Sesay holds an MCDC funded Ph.D. studentship. Dr Dianne Terlouw acknowledges support from the ACT consortium 
for the presented household prevalence surveys from Malawi. Prof Peter Diggle is supported
by the UK Medical Research Council (grant number G0902153).

\bibliographystyle{biometrika.bst}
\bibliography{article}

\begin{thebibliography}{23}
\expandafter\ifx\csname natexlab\endcsname\relax\def\natexlab#1{#1}\fi

\bibitem[{Christensen(2004)}]{mmle2004}
\textsc{Christensen, O.~F.} (2004).
\newblock {M}onte {C}arlo maximum likelihood in model-based geostatistics.
\newblock \textit{Journal of Computational and Graphical Statistics}
  \textbf{13}, 702--718.

\bibitem[{Christensen et~al.(2006)Christensen, Roberts \& Sk{\"o}ld}]{mcmc2006}
\textsc{Christensen, O.~F.}, \textsc{Roberts, G.~O.} \& \textsc{Sk{\"o}ld, M.}
  (2006).
\newblock Robust {M}arkov chain {M}onte {C}arlo methods for spatial generalized
  linear mixed models.
\newblock \textit{Journal of Computational and Graphical Statistics}
  \textbf{15}, 1--17.

\bibitem[{Christensen \& Waagepetersen(2002)}]{waag2002}
\textsc{Christensen, O.~F.} \& \textsc{Waagepetersen, R.~P.} (2002).
\newblock Bayesian prediction of spatial count data using generalized linear
  mixed models.
\newblock \textit{Biometrics} \textbf{58}, 280--286.

\bibitem[{Crainiceanu et~al.(2008)Crainiceanu, Diggle \&
  Rowlingson}]{crainiceanu2008}
\textsc{Crainiceanu, C.}, \textsc{Diggle, P.} \& \textsc{Rowlingson, B.}
  (2008).
\newblock Bivariate modelling and prediction of spatial variation in {\emph{
  \uppercase{l}oa loa}} prevalence in tropical {A}frica (with discussion).
\newblock \textit{Journal of the American Statistical Association}
  \textbf{103}, 21--43.

\bibitem[{Diggle et~al.(2010)Diggle, Menezes \& Su}]{prefer2010}
\textsc{Diggle, P.~J.}, \textsc{Menezes, R.} \& \textsc{Su, T.} (2010).
\newblock Geostatistical inference under preferential sampling.
\newblock \textit{Journal of the Royal Statistical Society, Series C}
  \textbf{59}, 191--232.

\bibitem[{Elliot \& Davis(2005)}]{cancer2005}
\textsc{Elliot, M.~R.} \& \textsc{Davis, W.~W.} (2005).
\newblock Obtaning risk factor prevalence estimates in small areas: combining
  data from two surveys.
\newblock \textit{Journal of the Royal Statistical Society, Series C}
  \textbf{54}, 595--609.

\bibitem[{Gahutu et~al.(2011)Gahutu, Steininger, Shyirambere, Zeile, Cwinya-Ay,
  Danquah, Larsen, Eggelte, Uwimana, Karema, Musemakweri, Harms \&
  Mockenhaupt}]{gahutu2011}
\textsc{Gahutu, J.-B.}, \textsc{Steininger, C.}, \textsc{Shyirambere, C.},
  \textsc{Zeile, I.}, \textsc{Cwinya-Ay, N.}, \textsc{Danquah, I.},
  \textsc{Larsen, C.}, \textsc{Eggelte, T.}, \textsc{Uwimana, A.},
  \textsc{Karema, C.}, \textsc{Musemakweri, A.}, \textsc{Harms, G.} \&
  \textsc{Mockenhaupt, F.} (2011).
\newblock Prevalence and risk factors of malaria among children in southern
  highland rwanda.
\newblock \textit{Malaria Journal} \textbf{10}, 134.

\bibitem[{Geyer(1994)}]{geyer1994}
\textsc{Geyer, C.~J.} (1994).
\newblock On the convergence of {M}onte {C}arlo maximum likelihood
  calculations.
\newblock \textit{Journal of the Royal Statistical Society, Series B}
  \textbf{56}, 261--274.

\bibitem[{Geyer(1996)}]{geyer1996}
\textsc{Geyer, C.~J.} (1996).
\newblock Estimation and optimization of functions.
\newblock In \textit{{M}arkov Chain {M}onte {C}arlo in Practice}, W.~Gilks,
  S.~Richardson \& D.~Spiegelhalter, eds. London: Chapman and Hall, pp.
  241--–258.

\bibitem[{Geyer(1999)}]{geyer1999}
\textsc{Geyer, C.~J.} (1999).
\newblock Likelihood inference for spatial point processes.
\newblock In \textit{Stochastic Geometry, Likelihood and Computation}, O.~E.
  Barndorff-Nielsen, W.~S.Kendall \& M.~N.~M. van Lieshout, eds. Boca Raton,
  FL: Chapman and Hall/CRC, pp. 79--–140.

\bibitem[{Geyer \& Thompson(1992)}]{geyer1992}
\textsc{Geyer, C.~J.} \& \textsc{Thompson, E.~A.} (1992).
\newblock Constrained {M}onte {C}arlo maximum likelihood for dependent data.
\newblock \textit{Journal of the Royal Statistical Society, Series B}
  \textbf{54}, 657--699.

\bibitem[{Hedt \& Pagano(2011)}]{hedt2011}
\textsc{Hedt, B.~L.} \& \textsc{Pagano, M.} (2011).
\newblock Health indicators: Eliminating bias from convenience sampling
  estimator.
\newblock \textit{Statistics in Medicine} \textbf{30}, 560--568.

\bibitem[{Lohr \& Rao(2006)}]{multframe2006}
\textsc{Lohr, S.~L.} \& \textsc{Rao, J. N.~K.} (2006).
\newblock Estimation in multiple-frame surveys.
\newblock \textit{Journal of the American Statistical Association}
  \textbf{101}, 1019--1030.

\bibitem[{Manzi et~al.(2011)Manzi, Spiegelhalter, Turner, Flowers \&
  Thompson}]{combsurv2011}
\textsc{Manzi, G.}, \textsc{Spiegelhalter, D.~J.}, \textsc{Turner, R.~M.},
  \textsc{Flowers, J.} \& \textsc{Thompson, S.~G.} (2011).
\newblock Modelling bias in combining small area prevalence estimates from
  multiple sruveys.
\newblock \textit{Journal of the Royal Statistical Society, Series A}
  \textbf{174}, 31--50.

\bibitem[{McCullagh \& Nelder(1989)}]{GLM1989}
\textsc{McCullagh, P.} \& \textsc{Nelder, J.} (1989).
\newblock \textit{Generalized Linear Models}.
\newblock Chapman and Hall, London, 2nd ed.

\bibitem[{Moriarity \& Scheuren(2001)}]{statmatch2001}
\textsc{Moriarity, C.} \& \textsc{Scheuren, F.} (2001).
\newblock Statistical matching: A paradigm for assesing the uncertainty in the
  procedure.
\newblock \textit{Journal of Official Statistics} \textbf{17}, 407--422.

\bibitem[{{R Core Team}(2012)}]{rsoftware}
\textsc{{R Core Team}} (2012).
\newblock \textit{R: A Language and Environment for Statistical Computing}.
\newblock R Foundation for Statistical Computing, Vienna, Austria.
\newblock {ISBN} 3-900051-07-0.

\bibitem[{Raghunathan et~al.(2007)Raghunathan, Xie, Schenker \&
  Parsons}]{raghu2007}
\textsc{Raghunathan, T.~E.}, \textsc{Xie, D.}, \textsc{Schenker, N.} \&
  \textsc{Parsons, V.~L.} (2007).
\newblock Combining information from two surveys to estimate county-level
  prevalence rates of cancer risk factors and screening.
\newblock \textit{Journal of the American Statistical Association}
  \textbf{102}, 474--486.

\bibitem[{Roca-Feltrer et~al.(2012)Roca-Feltrer, Lalloo, Phiri \&
  Terlouw}]{roca2012}
\textsc{Roca-Feltrer, A.}, \textsc{Lalloo, D.}, \textsc{Phiri, K.} \&
  \textsc{Terlouw, D.~J.} (2012).
\newblock \uppercase{R}olling \uppercase{M}alaria \uppercase{I}ndicator
  \uppercase{S}urveys (r\uppercase{MIS}): a potential district-level malaria
  monitoring and evaluation (\uppercase{M}\&\uppercase{E}) tool for programme
  managers.
\newblock \textit{American Journal of Tropical Medicine and Hygiene}
  \textbf{86}, 96--98.

\bibitem[{Turner et~al.(2009)Turner, Spiegelhalter, Smith \&
  Thompson}]{biasmod2009}
\textsc{Turner, R.~M.}, \textsc{Spiegelhalter, D.~J.}, \textsc{Smith, G. C.~S.}
  \& \textsc{Thompson, S.~G.} (2009).
\newblock Bias modelling in evidence synthesis.
\newblock \textit{Journal of the Royal Statistical Society, Series A}
  \textbf{172}, 21--47.

\bibitem[{Vyas \& Kumuranayake(2006)}]{vyas2006}
\textsc{Vyas, S.} \& \textsc{Kumuranayake, L.} (2006).
\newblock Constructing socio-economic status indices: how to use principal
  component analysis.
\newblock \textit{Health Policy Plan} \textbf{21}, 459--468.

\bibitem[{Wanji et~al.(2012)Wanji, Akotshi, Kankou, Nigo, Tepage, Ukety, Diggle
  \& Remme}]{wanji2012}
\textsc{Wanji, S.}, \textsc{Akotshi, D.}, \textsc{Kankou, J.}, \textsc{Nigo,
  M.}, \textsc{Tepage, F.}, \textsc{Ukety, T.}, \textsc{Diggle, P.} \&
  \textsc{Remme, J.} (2012).
\newblock The validation of the rapid assessment procedures for loiasis
  ({RAPLOA}) in the {D}emocratic {R}epublic of {C}ongo: health policy
  implications.
\newblock \textit{Parasites and Vectors} \textbf{5}, 25 {\tt
  doi:10.1186/1756--3305--5--25}.

\bibitem[{Zhang(2002)}]{em2002}
\textsc{Zhang, H.} (2002).
\newblock On estimation and prediction for spatial generalized linear mixed
  models.
\newblock \textit{Biometrics} \textbf{58}, 129--136.

\end{thebibliography}

\end{document}